\documentclass{emulateapj}
\usepackage{apjfonts}
\shorttitle{Optical and NIR Polarimetry of SN~2014J}
\shortauthors{Kawabata et al.}
\begin{document}
\title{
Optical and Near-Infrared Polarimetry of Highly Reddened Type Ia Supernova 2014J: 
Peculiar Properties of Dust in M82
}
\author{K. S. \textsc{Kawabata}\altaffilmark{1,2}, 
H. \textsc{Akitaya}\altaffilmark{1},
M. \textsc{Yamanaka}\altaffilmark{3,4},  
R. \textsc{Itoh}\altaffilmark{2,1},  
K. \textsc{Maeda}\altaffilmark{4,5},
Y. \textsc{Moritani}\altaffilmark{1}, 
T. \textsc{Ui}\altaffilmark{2},
M. \textsc{Kawabata}\altaffilmark{2}, 
K. \textsc{Mori}\altaffilmark{2},
D. \textsc{Nogami}\altaffilmark{4},
K. \textsc{Nomoto}\altaffilmark{5,13},
N. \textsc{Suzuki}\altaffilmark{5},
K. \textsc{Takaki}\altaffilmark{2},
M. \textsc{Tanaka}\altaffilmark{6},
I. \textsc{Ueno}\altaffilmark{2},
S. \textsc{Chiyonobu}\altaffilmark{2},
T. \textsc{Harao}\altaffilmark{2},
R. \textsc{Matsui}\altaffilmark{2},
H. \textsc{Miyamoto}\altaffilmark{2},
O. \textsc{Nagae}\altaffilmark{2},
A. \textsc{Nakashima}\altaffilmark{7},
H. \textsc{Nakaya}\altaffilmark{6},
Y. \textsc{Ohashi}\altaffilmark{2},
T. \textsc{Ohsugi}\altaffilmark{1},
T. \textsc{Komatsu}\altaffilmark{2},
K. \textsc{Sakimoto}\altaffilmark{2},
M. \textsc{Sasada}\altaffilmark{4},
H. \textsc{Sato}\altaffilmark{2},
H. \textsc{Tanaka}\altaffilmark{2},
T. \textsc{Urano}\altaffilmark{2},
T. \textsc{Yamashita}\altaffilmark{6},
M. \textsc{Yoshida}\altaffilmark{1,2},
A. \textsc{Arai}\altaffilmark{8},
N. \textsc{Ebisuda}\altaffilmark{2},
Y. \textsc{Fukazawa}\altaffilmark{2,1},
A. \textsc{Fukui}\altaffilmark{9},
O. \textsc{Hashimoto}\altaffilmark{10},
S. \textsc{Honda}\altaffilmark{8,10},
H. \textsc{Izumiura}\altaffilmark{9},
Y. \textsc{Kanda}\altaffilmark{2},
K. \textsc{Kawaguchi}\altaffilmark{2},
N. \textsc{Kawai}\altaffilmark{11},
D. \textsc{Kuroda}\altaffilmark{9},
K. \textsc{Masumoto}\altaffilmark{12},
K. \textsc{Matsumoto}\altaffilmark{12},
T. \textsc{Nakaoka}\altaffilmark{2},
K. \textsc{Takata}\altaffilmark{2},
M. \textsc{Uemura}\altaffilmark{1,2},
and K. \textsc{Yanagisawa}\altaffilmark{9}
}

\altaffiltext{1}{Hiroshima Astrophysical Science Center, Hiroshima University, Kagamiyama, 
Higashi-Hiroshima, Hiroshima 739-8526, Japan; kawabtkj@hiroshima-u.ac.jp} 
\altaffiltext{2}{Department of Physical Science, Hiroshima University, Kagamiyama, 
Higashi-Hiroshima 739-8526, Japan} 
\altaffiltext{3}{Department of Physics, Faculty of Science and Engineering, Konan University, 
Okamoto, Kobe, Hyogo 658-8501, Japan}
\altaffiltext{4}{Department of Astronomy, Graduate School of Science,
Kyoto University, Sakyo-ku, Kyoto 606-8502, Japan}
\altaffiltext{5}{Kavli Institute for the Physics and Mathematics of the Universe (WPI), 
The University of Tokyo, Kashiwa, Chiba 277-8583, Japan}
\altaffiltext{6}{National Astronomical Observatory of Japan,  Osawa, Mitaka, Tokyo 181-8588, Japan}
\altaffiltext{7}{Nagoya City Science Museum, Sakae, Naka-ku, Nagoya 460-0008, Japan}
\altaffiltext{8}{Nishi-Harima Astronomical Observatory, Center for Astronomy, University of Hyogo,
Nishigaichi, Sayo-cho, Sayo, Hyogo 679-5313, Japan}
\altaffiltext{9}{Okayama Astrophysical Observatory, NAOJ, Honjo, Kamogata-cho Asauchi, Okayama 719-0232, Japan}
\altaffiltext{10}{Gunma Astronomical Observatory, Takayama, Gunma 377-0702}
\altaffiltext{11}{Department of Physics, Tokyo Institute of Technology,Ookayama, Meguro-ku, Tokyo 152-8551, Japan}
\altaffiltext{12}{Institute of Astronomy, Osaka Kyoiku University, Kashiwara, Osaka 582-8582, Japan}
\altaffiltext{13}{Hamamatsu Professor}
\begin{abstract}
We presented optical and near-infrared multi-band linear polarimetry of the 
highly reddened Type Ia SN~2014J appeared in M82.
SN~2014J exhibits large polarization at shorter wavelengths,
e.g., $4.8$\% in $B$ band, and the polarization decreases rapidly
at longer wavelengths, with the position angle of the polarization
remaining at approximately $40^{\circ}$ over the observed wavelength range.
These polarimetric properties suggest that the observed polarization is likely 
to be caused predominantly by the interstellar dust within M82.
Further analysis shows that the polarization peaks at a wavelengths much shorter
than those obtained for the Galactic dust. 
The wavelength dependence of the polarization can be
better described by an inverse power law rather than by Serkowski law 
for Galactic interstellar polarization.
These suggests that the nature of the dust in M82 may be different 
from that in our Galaxy, with polarizing dust grains having a mean radius of 
$<0.1\ \mu$m .
\end{abstract}


\section{Introduction}
\label{sec:intro}

The homogeneity in photometric properties of normal Type Ia supernovae (SNe Ia)
is expected to be related to common physical properties during the onset of 
thermonuclear explosions in the progenitor white dwarfs with a mass close to
Chandrasekhar's limiting mass \cite[][for review]{hil00}. 
The continuum light from normal SNe Ia is intrinsically weakly
polarized ($p\lesssim 0.3$\%), although the absorption features, including 
\ion{Si}{2} 6355 and \ion{Ca}{2} IR triplets, are often polarized by $0.5$--$1.5$\%
\citep{wan96,wan97,wan03,leo05,wan06,wan08,cho08,zel13,mau13}.
In fact, a \ion{Si}{2} 6355 absorption line with a polarization of $p=1.5$\%
and an equivalent width of $0.011\ \mu$m (for SN~2014J near the maximum) 
gives an additional polarization of only $\Delta p=0.13$\% for typical
$R_{C}$ band polarimetry ($\Delta \lambda=0.13\ \mu$m).
This practically allows us to use SNe Ia as a unique bright
unpolarized-light source within distant galaxies for broadband polarimetry.
Thus, a SN Ia has the potential to project the interstellar polarization (ISP)
along the line of sight inside the host galaxy when subject to 
a substantial amount of interstellar reddening, as is commonly seen in 
our Galaxy \citep[e.g.,][]{whi03}.

SN~2014J is the closest SN Ia in this quarter century.
It was discovered in M82 (at a distance $\sim 3.9\pm 0.4$ Mpc; \citealt{sak99}) 
on 2014 Jan 21.81 (UT dates are used throughout this Letter)
at a magnitude of $R=10.99\pm 0.03$ mag \citep{fos14,goo14}, 
probably a week after the explosion \citep{zhe14}.
The apparent brightness provides us with an opportunity for various studies,
including constraining early light curve model
\citep{goo14, zhe14}, studying progenitor systems \citep{kel14}, and
evaluating properties of extragalactic interstellar/circumstellar 
media \citep{ama14,fol14, wel14}.
In addition, this SN provides a rare opportunity to probe the ISP within 
the starburst galaxy M82 because it suffers significant reddening from the
host galaxy ($E_{B-V}^{\rm host}\sim 1.3$ mag).
Prior to SN~2014J, there were only three reddened SNe Ia ($E_{B-V}^{\rm host}\geq 0.5$ mag)
for which the wavelength dependence of optical polarization had been measured,
which are SN 1986G in the peculiar giant S0 galaxy NGC 5128 (Cen A) 
(\citealt{hou87}; $E_{B-V}^{\rm host}\simeq 1.6$ mag and $d\simeq 4$ Mpc),
SN 2006X in the Virgo Cluster spiral galaxy NGC 4321 (\citealt{pat09}; 
$E_{B-V}^{host}=1.5$--$1.7$ mag and $d\simeq 16$ Mpc),
and SN 2008fp in the peculiar spiral galaxy ESO 428-G14 (\citealt{cox14}; 
$E_{B-V}^{host}=0.6\pm 0.1$ mag and $d\sim 26$ Mpc).

The Galactic ISP at ultraviolet (UV) to near-infrared (NIR) wavebands can be
approximated by Serkowski law (\citealt{ser75}), a smooth function of wavelength
given by
\begin{equation}
 p(\lambda) = p_{\rm max} \exp\left[ -K \ln^{2}\left( \frac{\lambda_{\rm max}}{\lambda}
 \right) \right] \mbox{ ,}
\end{equation}
where $p_{\rm max}$ is 
the peak polarization degree occurring at wavelength $\lambda_{\rm max}$ and 
$K$ is a parameter describing the width of the peak.
The polarization observed for SNe 1986G and 2006X can be described by
Serkowski law at optical wavelengths; however, the derived parameters are
peculiar, i.e., the wavelengths $\lambda_{\rm max}=0.43\pm 0.01 \mu$m (SN 1986G) and 
$\lambda_{\rm max}=0.35\pm 0.01 \mu$m (SN 2006X) are significantly shorter
than the Galactic value ($0.54\pm 0.06 \mu$m; \citealt{vrb81}), 
and $K=1.3\pm 0.1$ (SN 2006X) is not consistent with the value expected from the
Wilking law, i.e., $K=(1.66\pm 0.09)\lambda_{\rm max}(\mu\mbox{m})+(0.01\pm 0.05)$,
for the Galactic ISP (\citealt{wil80,whi92}).
SN 2008fp exhibits interstellar polarization similar to the downscaled one 
of SN 2006X \citep{cox14}.
For the Galactic ISP, this $\lambda_{\rm max}$--$K$ correlation may be 
interpreted as a narrowing in the size distribution with grain growth \citep[e.g.,][]{whi03}.
The shorter $\lambda_{\rm max}$ with these SNe suggests 
that the size of the dust grains polarizing light in the host galaxies is, 
on average,
smaller than those in the Milkyway. This is also consistent with the
smaller values of the total-to-selective extinction ratio, i.e.,
$R_{V}=A_{V}/E_{B-V}\simeq 1.3$--$2.6$, 
obtained for some reddened SNe Ia \citep[][and references therein]{phi13}.
If the empirical relation of Galactic ISP, 
$R_{V}=(5.6\pm 0.3)\lambda_{\rm max}(\mu\mbox{m})$ \citep{ser75,whi03},
still holds for small $\lambda_{\rm max}$, the observed values of $R_{V}\simeq 1.3$--$2.6$
corresponds to $\lambda_{\rm max}\simeq 0.23$--$0.46 \mu$m, which is 
comparable with $\lambda_{\rm max}$ observed in SNe 1986G and 2006X. 

In this Letter, we report our $BVR_{C}I_{C}JHK_{s}$ polarimetry 
of SN~2014J before and after the maximum light, along with our photometric
and spectroscopic observations.
Such multi-band polarimetry including NIR bands for reddened SNe 
Ia is still quite rare, and therefore this SN may provide us a
valuable information on the interstellar and/or circumstellar media
along the line of sight toward SN~2014J within M82.

\section{Observations and Reduction}
\label{obs_red}

We performed imaging polarimetry of SN~2014J using Hiroshima One-shot 
Wide-field Polarimeter (HOWPol; \citealt{kaw08}) on 2014 Jan 22.4 
($t=-11.0$ days relative to the $B$-band maximum light; see \S \ref{subsec:phot_spec})
in $VR_{C}I_{C}$ bands and Hiroshima Optical and Near IR camera
(HONIR; \citealt{aki14}) in $BVR_{C}I_{C}JHK_{s}$ bands
on Jan 27.7 ($-5.7$ days), Feb 16.5 ($+14.1$ days), 25.6 ($+23.2$ days) and 
Mar 7.8 ($+33.4$ days).
HOWPol employs a wedged double Wollaston prism, and is attached to the Nasmyth focus of the 
1.5~m Kanata telescope at Higashi-Hiroshima Observatory.
HONIR uses a cooled LiYF$_{4}$ Wollaston prism and is attached to the Cassegrain
focus of the same telescope.
Each observation consisted of a sequence of exposures
at four position angles (PAs) of the achromatic half-wave plates, 
$0\fdg 0$, $22\fdg 5$, $45\fdg 0$, 
and $67\fdg 5$ for the HOWPol and the last HONIR observations, and
at four PAs of the instrumental rotator at the Cassegrain focus of 
the telescope, $0^{\circ}$, $90^{\circ}$, $45^{\circ}$ and $135^{\circ}$,
for the first three HONIR observations.
These data were calibrated using observations of unpolarized 
(HD~94851, HD~98281) and
polarized standard stars (HD~30168, HD~150193, HDE~283701, Cyg~OB~2 \#11;
\citealt{tur90,whi92}), including measurements 
through a fully-polarizing filter or a wire grid.
Using this procedure, the instrumental polarization ($p\lesssim 0.2$\% in
HONIR and $p\simeq 3$--$4$\% in HOWPol) was vectorially removed.

In addition, we obtained photometry using HOWPol ($BVR_{C}I_{C}$), and HONIR
($BVR_{C}I_{C}JHK_{s}$) attached to the 1.5-m Kanata telescope, 
with MITSuME \citep{kot05} ($g'R_{C}I_{C}$) attached 
to the 0.5~m telescope at Okayama Astrophysical Observatory (OAO) of 
National Astronomical Observatory of Japan, with a Peltier-cooled CCD
($BVR_{C}I_{C}$) attached to the 0.51~m telescope at Osaka Kyoiku University (OKU), and with 
ISLE \citep{yan08} ($JHK_{s}$) attached to the 1.88~m telescope at OAO, respectively.
The magnitude in each band was determined relative to the nearby 
comparison star, BD+70$^{\circ}$587, which was flux-calibrated in $BVR_{C}I_{C}$ bands
using Landolt field stars \citep{lan92} on a photometric night.
For NIR photometry, we used $JHK_{s}$ magnitudes of the same star in
2MASS Second Incremental Release Point Source Catalogue.
We also collected low-resolution  spectra with HOWPol (0.41--0.94 $\mu$m, 
$R=\lambda/\Delta\lambda\simeq 400$) and HONIR (0.5--2.3 $\mu$m, 
$R\simeq 450$--$600$) on the 1.5-m Kanata telescope.
The flux was calibrated using observations of spectrophotometric
standard stars obtained on the same nights.

Because SN~2014J is superimposed within the bright region of M82 and
we cannot perform template subtraction in image reduction, 
the obtained flux should be more or less contaminated by the 
inhomogeneity and irregularity of the surface brightness of M82.
However, the SN itself is sufficiently bright, and 
the polarizations obtained during the period from $t=-11$ days to
$t=+33$ days from the maximum light should suffer minor effects 
from the galaxy light.

\section{Results}

\subsection{Photometric and Spectroscopic Properties}
\label{subsec:phot_spec}

\begin{figure}
  \begin{center}
    \begin{tabular}{c}
      \resizebox{130mm}{!}{\includegraphics{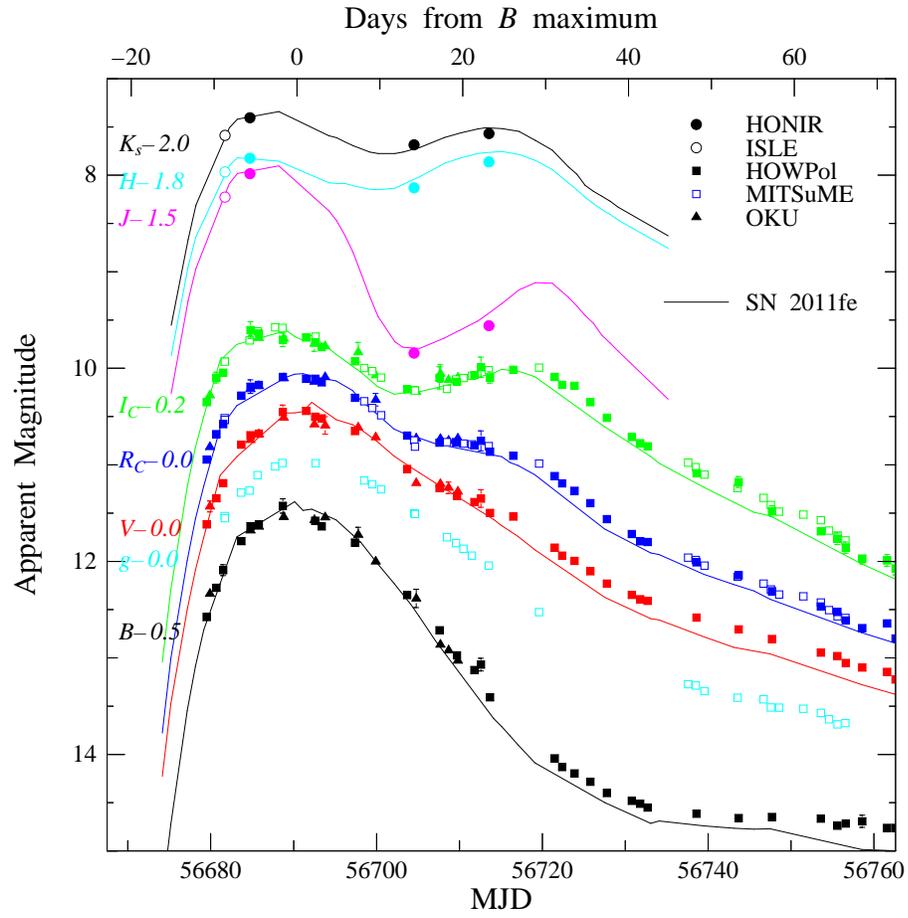}} \\
    \end{tabular}
    \caption{Multi-band light curves (LCs) of SN~2014J. 
      We also plot LCs of normal SN Ia 2011fe 
      ($\Delta m_{B}(15)=1.21\pm 0.03$ mag;
      \citealt{ric12,mat12}) 
      for comparison, which are shifted to match the peak time and magnitudes.
    }
    \label{fig:fig1}
    \end{center}
\end{figure}

Figure \ref{fig:fig1} shows the obtained multi-band light curves (LCs).
The apparent maximum magnitudes in $BV$ bands are found to be 
$B_{\rm max}=11.99\pm 0.05$ mag on MJD 56690.4$\pm 0.5$ (Feb 2.4$\pm 0.5$) 
and $V_{\rm max}=10.44\pm 0.03$ mag on MJD 56691.7$\pm 0.5$, respectively, 
as derived by a polynomial fit to the observed data around the maximum light.
We also derived the observed $B$-band magnitude decline rate
$\Delta m_{15}(B)=1.02\pm 0.05$ mag.
For the extinction toward SN~2014J, \citet{ama14} estimated 
the total reddening of $E_{B-V}^{\rm total}=1.37\pm 0.03$ mag and 
$R_{V}^{\rm total}=1.4\pm 0.1$ based on analysis of near-maximum-light 
spectral energy distribution from UV to NIR wavebands.
This reddening is apparently dominated by the host galaxy component 
$E_{B-V}^{host}$, because the IR Dustmap suggests that the Galactic 
component is only $E_{B-V}^{\rm MW}=0.14$ mag \citep{sch11}.
We corrected for the extinction using the 
$E_{B-V}^{\rm total}$ and $R_{V}^{\rm total}$ values and 
the parameterized extinction curve \citep{car89}.
The absolute magnitudes of $M_{B,{\rm max}}=-19.26\pm 0.26$ mag
and $M_{V,{\rm max}}=-19.42\pm 0.25$ mag, as well as its color, are 
consistent with the empirical relations with $\Delta m_{B}(15)$ 
within errors \citep[e.g.,][]{phi99}, suggesting the photometric
behavior in SN~2014J is not anomalous.
We set the time of the $B$ band maximum as $t=0$ days,
which is $18.7\pm 0.5$ days after the epoch of the estimated first light 
\citep{zhe14,goo14}.

\begin{figure}
  \begin{center}
    \begin{tabular}{c}
      \resizebox{130mm}{!}{\includegraphics{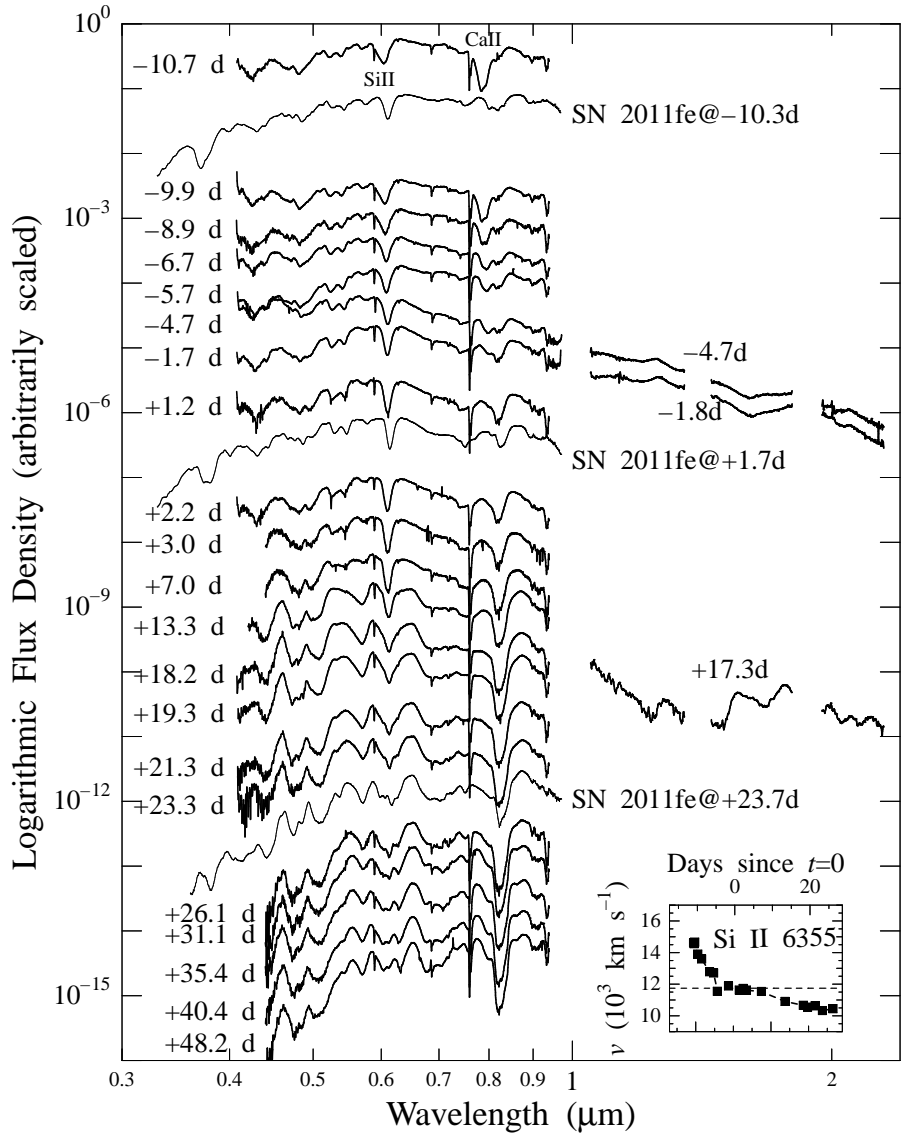}} \\
    \end{tabular}
    \caption{Spectral evolution of SN~2014J. The epoch of each spectrum is indicated
      in the panel.  
      For comparison, we also plot optical spectra of SN 2011fe at three epochs 
      \citep{per13}, reddened with $E_{B-V}=1.37$ and $R_{V}=1.4$ to match 
      those of SN~2014J.
      The inset panel shows the evolution of the line velocity 
      of \ion{Si}{2} 6355 and the horizontal dashed line shows the velocity 
      at around $t=0$ days (see \ref{subsec:phot_spec}).
    }
    \label{fig:fig2}
    \end{center}
\end{figure}

Figure \ref{fig:fig2} shows a time series of spectra from 
$t=-11$ days to $t=+48$ days.
Compared with the normal SN Ia 2011fe, SN~2014J is characterized by 
the absence of spectral features due to \ion{C}{2} 6580 and \ion{O}{1} 7774,
as well as the existence of high-velocity components in
\ion{Ca}{2} IR triplet ($\gtrsim 20,000$ km s$^{-1}$) 
during the earliest phase $t\lesssim -5$ days, as 
pointed out by \citet{goo14}.
The line velocity and equivalent width of \ion{Si}{2} 6355
around maximum ($-2<t<2$ days) are $-11,750\pm 300$ km s$^{-1}$ 
(Figure \ref{fig:fig2} inset panel) and $110\pm 5$ \AA, 
respectively, which are marginal between those of `Normal' 
and `HV' SNe \citep{wan09}.
However, the relation between $\Delta m_{15}$-corrected $M_{V,{\rm max}}$ 
and $E_{B-V}^{\rm host}$ yielded in SN~2014J ($\sim -17.9$ mag for
$M_{V,{\rm max}}$ and $\sim 1.23$ mag for $E_{B-V}^{\rm host}$)
is apparently consistent with the
branch of HV group ($R_{V}\sim 1.6$; \citealt{wan09}).
The nearly constant absorption strength (up to $\sim 3$ months
after discovery) of the interstellar 
\ion{Na}{1} D lines and diffuse interstellar bands
\citep[e.g.,][]{wel14} indicate that the dust responsible
for the extinction towards SN~2014J is located at a site
moderately separated from the progenitor ($\gtrsim 2\times 10^{16}$ cm).

\subsection{Polarimetric Properties}
\label{subsec:polari}

\begin{figure}
  \begin{center}
    \begin{tabular}{c}
      \resizebox{120mm}{!}{\includegraphics{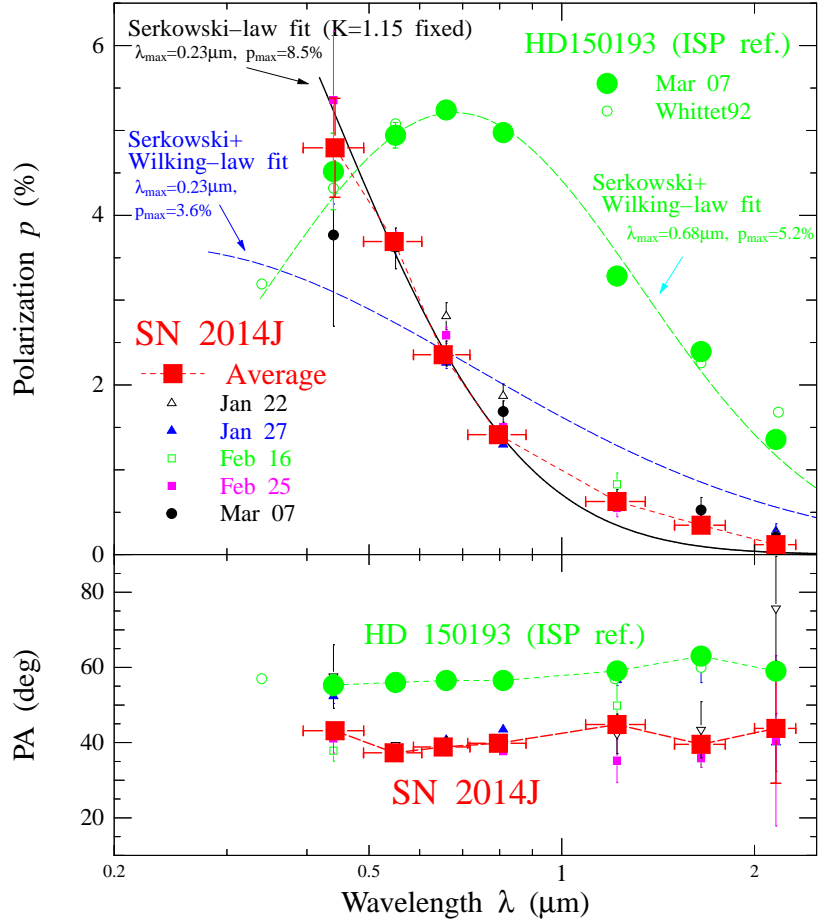}} \\
    \end{tabular}
    \caption{Result of our multi-band polarimetry. 
      The upper panel shows the degree of 
      polarization and the lower panel shows the position angle on the 
      projected sky. The red squares denote the polarization of SN~2014J
      averaged over all five/four nights (Table \ref{tbl:pol_sn14j}),  
      and the small symbols show the individual nightly data, as indicated.
      The curves in the panel show the empirical laws of the Galactic ISP, 
      Serkowski law (black line), and that with Wilking law (blue dashed line), 
      fitted to the averaged polarization data at optical wavelengths 
      ($\lambda<1\ \mu$m). 
      For comparison, we plot the observed/cataloged polarization 
      (green circles/dots) of the strongly-polarized standard star, 
      HD~150193 \citep{whi92}, as a typical $p(\lambda)$ curve of Galactic ISP.
      It is clear that the strong wavelength dependence of SN~2014J is not
      readily explained by the Wilking law.
     }
    \label{fig:fig3}
    \end{center}
\end{figure}

The observed polarization is shown in Figure \ref{fig:fig3}.
The polarization is relatively strong in blue bands, e.g., reaching $\sim 4.8$\% in $B$ band, 
and it decreases rapidly with wavelength, while the polarization
PA is approximately constant at around $40^{\circ}$.
In NIR bands, the polarization is less significant ($p\lesssim 1$\%); however, 
it is likely that the same polarization component still dominates because of 
having almost the same PA as the optical bands.
There is no significant temporal variation in the polarization measured during
the period from $t=-11$ days to $t=+33$ days from the maximum light, and 
the polarization in optical bands appears to be consistent with the result of  
spectropolarimetry covering wavelengths from 380 to 880 nm \citep{pat14}. Hereafter, 
we discuss only the averaged polarization (Table \ref{tbl:pol_sn14j}).

The large polarization measured for SN~2014J suggests that it is predominantly produced 
within M82, because the Galactic ISP is, at most, 0.18\% according to the measurements of
six Galactic stars in the vicinity of SN~2014J within $10^{\circ}$ of the 
all-sky polarization map \citep{hei00}.
Furthermore, the almost constant polarization during the period of 
our observation would exclude the possibility that it is originated in the
close proximity to the progenitor. This, together with the unchanged 
interstellar absorption lines, favors that that the significant polarization 
of SN~2014J is caused by dust grains at a site remote from the SN.
It has been suggested that multiple scattering due to circumstellar 
(CS) dust may account for half of the total extinction \citep{fol14}.
However, we argue that the
CS dust could not be the principal origin of the observed polarization because
multiple scattering would effectively depolarize the light and the resulting
continuum polarization would also show significant changes with time
(see also \citealt{pat14}).
In the optical image of M82 \citep[e.g.,][]{ohy02}, PA of $\sim 40^{\circ}$
seems to align with the direction of the local dust lanes around the SN position,
which further strengthens the argument that the polarization is unrelated
to the CS matter.

Figure \ref{fig:fig3} shows that the polarization peak
appears outside the wavelength range of our observations, 
i.e., $\lambda_{\rm max}\lesssim  0.4 \mu$m.
This $\lambda_{\rm max}$ is considerably smaller than the typical value 
determined from the Galactic ISP.
For comparison, we also plotted the polarization curves determined from the
Serkowski law with/without the Wilking law in Figure \ref{fig:fig3}.
For the Serkowski law, we fitted it with a constant $K\equiv 1.15$, 
a typical one for Galactic ISP \citep{ser75},
because the fitted parameters do not converge in case of free $K$.
We cannot obtain any good fit with the Wilking law. In general, with 
the Wilking law, a short $\lambda_{\rm max}$ leads to small $K$ (corresponding to
broader peak in $p(\lambda)$ curve); however, the observed steep gradient of
$p(\lambda)$ requires a large $K$ value.
The Wilking law also fails to describe the continuum polarization 
measured for SN 2006X \citep{pat09}.
In addition, we find that 
$K=1.13\pm 0.05$ and $\lambda_{\rm max}=0.43\pm 0.01 \mu$m
derived for the SN 1986G data \citep{hou87} do not satisfy the Wilking law.
The fact that 5 out of 105 Galactic reddened stars show 
considerable ISP with $\lambda_{\rm max}< 0.4\ \mu$m and the Wilking law
holds for 4 of the 5 stars within the errors \citep{whi92} suggests that
the failure of the Wilking law to describe the data may be common for 
highly reddened SNe Ia and the dust properties of the host galaxies 
may differ from those of the Milkyway.

\begin{figure}
  \begin{center}
    \begin{tabular}{c}
      \resizebox{130mm}{!}{\includegraphics{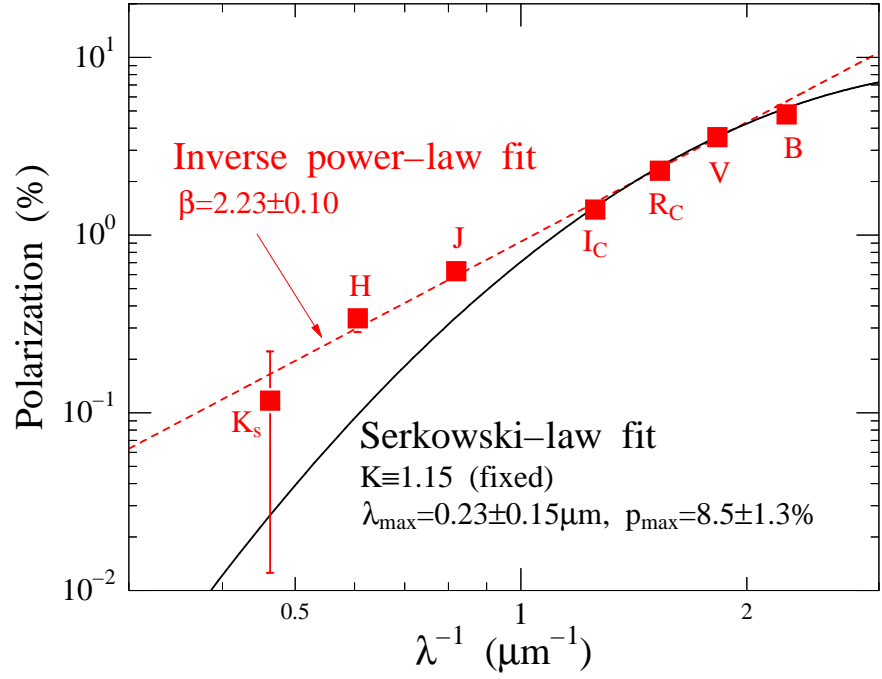}} \\
    \end{tabular}
    \caption{Polarization curve of SN~2014J as a function of the inverse wavelength
      plotted on logarithmic axes. 
      The red squares and the black line are the same in Fig. \ref{fig:fig3}.
      The red dashed straight line shows a power-law fit to the polarization data 
      reported in this paper, except for $B$-band data.
      The wavelength dependence of polarization at regions with 
      $\lambda > 0.5\ \mu$m can be better explained by 
      $\propto \lambda^{-(2.23\pm 0.10)}$, rather than Serkowski law
      (black line).
    }
    \label{fig:fig4}
    \end{center}
\end{figure}

Using only Serkowski law, we note that there is a systematic difference in 
the polarization of $\Delta p=0.2$--$0.3$\% between the observed polarization 
and the fitted curve at longer wavelengths ($\gtrsim 1\ \mu$m, Figure \ref{fig:fig3}).
This difference may be explained by considering
an analog of the `IR polarization excess' found in the 
Galactic ISP at longer wavelengths
($\gtrsim 2\ \mu$m), which is characterized by an inverse power-law, i.e.,
$p(\lambda)$ \citep[e.g.,][]{nag90}.
The wavelength dependence of polarization of SN~2014J at 
$\gtrsim 0.5\ \mu$m can be well described by $p(\lambda )=p_{1}\lambda^{-\beta}$ 
with the index of $\beta=2.23\pm 0.10$ (Figure \ref{fig:fig4}).
It should be noted that the polarization closely follows a power-law 
dependence even at the optical wavelengths for SN~2014J.
For Galactic reddened stars,
the index $\beta$ is typically in a relatively narrow range of 1.5--2.0
and is uncorrelated with the wavelength dependence of optical polarization, 
e.g., $\lambda_{\rm max}$ \citep{mar92,whi03}.
The index $\beta$ obtained for the ISP in M82 appears slightly steeper than
that for the Galactic ISP.
This may be related to the failure of the Wilking law, because the polarization
peak observed in SN~2014J is clearly sharper than that expected for a
Galactic ISP with a similar $\lambda_{\rm max}$ (see Figure \ref{fig:fig3}).

\section{Discussion}
\label{sec:discussion}

As described above, SN~2014J is a highly reddened SN Ia,
similar to SNe 1986G, 2006X and 2008fp.
The blue continuum of these SNe Ia all exhibit significant polarization, 
which is atypical of a Galactic ISP.
To a first-order approximation, a small $\lambda_{\rm max}$ corresponds
to small mean size of the dust grains causing the observed polarization.
This can be explained using Mie theory for dielectric cylinders,
e.g., $\lambda_{\rm max}\sim 2\pi a_{\rm eff} (n-1)$, where $a_{\rm eff}$ is the
effective radius and
$n$ is refractive index of the cylindrical grain \citep[e.g.,][]{whi03}.
Assuming $\lambda_{\rm max}\lesssim 0.4\ \mu$m and
$n=1.6$ (appropriate for silicates), then $a_{\rm eff}$ of 
the polarizing grains should be $\lesssim 0.11\ \mu$m.
Although a small $\lambda_{\rm max}$ (and thus a small $a_{\rm eff}$) could be the 
result of a failure of alignment/asphericity only of larger grains, the small 
$R_{V}$ inferred for SN~2014J suggests that the effect is not significant 
and the grain size should be intrinsically small.

It is not known whether the empirical relation $R_{V}=(5.6\pm 0.3)\lambda_{\rm max}$($\mu$m) 
determined for the Galactic ISP \citep{ser75,whi03} holds in the host galaxies of 
these highly reddened SNe Ia; however, interestingly, it has been shown that 
$R_{V}$ is small for these SNe, e.g., $2.57^{+0.23}_{-0.21}$
for SN 1986G and $1.31^{+0.08}_{-0.10}$ for SN 2006X
\citep[Table 2 in][]{phi13} and the $\lambda_{\rm max}=0.43\pm 0.01\ \mu$m obtained
for SN 1986G \citep{hou87} satisfies the empirical relation within the errors.
For SN 2006X, $R_{V}=1.31$ is much smaller than that expected from
$\lambda_{\rm max}\sim 0.35\ \mu$m (i.e., $R_{V}=2.0\pm 0.1$).
\citet{wanx08} derived a slightly larger value of $R_{V}\sim 1.5$ from optical and NIR 
photometry of SN 2006X; however, this is still smaller than the expected value.
It should be noted that $\lambda_{\rm max}$ is somewhat ambiguous because the 
observations did not cover the wavelengths shorter than 0.35\ $\mu$m
\citep{pat09}, which makes it difficult to completely rule out the applicability 
of the empirical relation.
For SN~2014J, $R_{V}^{\rm total}=1.4\pm 0.1$ \citep{ama14}, which 
leads to $\lambda_{\rm max}=0.25\pm 0.02\ \mu$m, which is 
speculatively consistent with the result of our fit to Serkowski law
with a constant $K=1.15$.
Therefore, although the Wilking law (i.e., $\lambda_{\rm max}$--$K$ relation) 
no longer appears to be valid for extragalactic ISPs (see \S \ref{subsec:polari}), 
the positive correlation of $\lambda_{\rm max}$--$R_{V}$ may hold 
even for an ISP with $\lambda_{\rm max}<0.4\ \mu$m, 
at least, in part of the extragalactic environment.
If this is indeed the case, the smaller values of $R_{V}$ seen in some moderately to 
highly reddened SNe Ia (e.g., \citealt{phi13}) suggests that the ISP in the host galaxies
has a smaller $\lambda_{\rm max}$, and hence a smaller $a_{\rm eff}$. 
The failure of the data to follow the Wilking law is harder to be explain
using grain sizes only; a qualitative difference in the size distribution
(and possibly in composition, shape, and degree of alignment) would be required for 
the dust grains between our Galaxy and those host galaxies of SNe Ia.
A possible explanation for the atypical dust seen in the host galaxies 
of some SNe Ia is that the interstellar dust within our Galaxy is not in fact typical.
The understanding of physical properties of extragalactic dust grains remains incomplete.
Extensive data on the UV and optical polarimetry for SNe Ia
may therefore be crucial for further understanding of the properties of extragalactic dust.

\begin{acknowledgements}
This work was supported by JSPS Research Fellowships for Young
Scientists (KT), by the Grant-in-Aid for Scientific Research from JSPS 
(23340048,23740141,26287031,26800100), and by Optical and NIR Astronomy Inter-University
Cooperation Program, OISTER, and by WPI Initiative, from the Ministry of 
Education, Culture, Sports, Science, and Technology in Japan.
\end{acknowledgements}

\begin{deluxetable}{cccccc}
  \tabletypesize{\scriptsize}
  \tablecaption{Measured polarization of SN~2014J\label{tbl:pol_sn14j}}
  \tablewidth{0pt}
  \tablehead{
   \colhead{Band} & 
   \colhead{$p$ (\%)\tablenotemark{a}} &
   \colhead{PA\tablenotemark{a}} &
   \colhead{Band} & 
   \colhead{$p$ (\%)\tablenotemark{a}} &
   \colhead{PA\tablenotemark{a}}}
  \startdata
  $B$     & 4.79$\pm$0.58 & $43\fdg 2\pm 1\fdg 8$ &  $J$     & 0.63$\pm$0.06 & $44\fdg 8\pm 3\fdg 3$  \\
  $V$     & 3.69$\pm$0.14 & $37\fdg 3\pm 1\fdg 1$ &  $H$     & 0.35$\pm$0.06 & $39\fdg 5\pm 2\fdg 5$ \\
  $R_{C}$ & 2.36$\pm$0.06 & $38\fdg 8\pm 0\fdg 9$ &  $K_{s}$  & 0.12$\pm$0.10 & $43\fdg 8\pm 14\fdg 6$ \\
  $I_{C}$ & 1.42$\pm$0.08 & $39\fdg 8\pm 1\fdg 4$ &          &               &               \\
  \enddata
  \tablenotetext{a}{Averaged polarization and the position angle are shown
    (see \S\ref{subsec:polari}).
    In $VRI$ bands the data are weighted means over five nights from 
    $t=-11$ days through $t=+33$ days, and in other bands they are over four nights
    from $t=-6$ days through $+33$ days. The error is predominantly due
    either to the observational error ($\sigma$) in $R_{C}I_{C}JHK_{s}$ bands 
    or the uncertainty of the polarimetric calibration (instrumental
    polarization/depolarization) in $BV$ bands.
  }
\end{deluxetable}

\vspace*{-0.5ex}

\begin{thebibliography}{10}

\bibitem[Akitaya et~al.(2014)]{aki14}
Akitaya, H., Moritani, Y., Ui, T., et al. 2014,
Proc. SPIE, 9147, 91474O

\bibitem[Amanullah et al.(2014)]{ama14}
Amanullah, R., Goobar, A., Johansson, J., et al. 2014, \apjl, 788, L21
  
\bibitem[Cardelli, Clayton, \& Mathis(1989)]{car89}
Cardelli, J. A., Clayton, G. C., \& Mathis, J. S. 1989, \apj, 345, 245
  
\bibitem[Chornock \& Filippenko(2008)]{cho08}
Chornock, R., \& Filippenko, A. V. 2008, \aj, 136, 2227

\bibitem[Cox \& Patat(2014)]{cox14}
Cox, N. L. J., \& Patat, F. 2014, \aap, 565, A61
  
\bibitem[Fossey et al.(2014)]{fos14}
Fossey, S., Cooke, B., Pollack, G., Wilde, M., \& Wright, T. 2014, CBET, 3792, 1

\bibitem[Foley et al.(2014)]{fol14}
Foley, R. J., Fox, O., McCully, C., Phillips, M. M., Sand, D. J., et al. 2014, \mnras, submitted (arXiv: 1405.3677)

\bibitem[Goobar et al.(2014)]{goo14}
Goobar, A., Johansson, J., Amanullah, R., et al. 2014, \apjl, 784, L12

\bibitem[Heiles(2000)]{hei00} Heiles, C. 2000, \aj, 119, 923

\bibitem[Hillebrandt \& Niemeyer(2000)]{hil00}
Hillebrandt, W., \& Niemeyer, J. C. 2000, \araa, 38, 191

\bibitem[Hough et al.(1987)]{hou87}
Hough, J. H., Bailey, J. A., Rouse, M. F., \& Whittet, D. C. B. 1987,\mnras, 227, 1

\bibitem[Hutton et al.(2014)]{hut14}
Hutton, S., Ferreras, I., Wu, K., et al. 2014, \mnras, 440, 150

\bibitem[Kawabata et~al.(2008)]{kaw08}
Kawabata, K. S., Nagae, O., Chiyonobu, S., et al. 2008, Proc. SPIE, 7014, 70144L

\bibitem[Kelly et al.(2014)]{kel14}
Kelly, P. L., Fox, O. D., Filippenko, A. V., et al. 2014, \apj, 790, 3

\bibitem[Kotani et~al.(2005)]{kot05}
Kotani, T., Kawai, N., Yanagisawa, K., et al. 2005, NCimC, 28, 755

\bibitem[Leonard et al.(2005)]{leo05}
Leonard, D. C., Li, W., Filippenko, A. V., Foley, R. J., \& Chornock, R. 2005,
\apj, 632, 450

\bibitem[Landolt(1992)]{lan92}
Landolt, A. U. 1992, \aj, 104, 340

\bibitem[Marion et al.(2014)]{mar14}
Marion, G. H., Sand, D. J., Hsiao, E. Y., Banerjee, D. P. K., Valenti, S., et al., \apj, submitted (arXiv: 1405.3970)

\bibitem[Martin et al.(1992)]{mar92}
Martin, P. G., Adamson, A. J., Whittet, D. C. B., et al. 1992, \apj, 392, 691

\bibitem[Matheson et al.(2012)]{mat12}
Matheson, T, Joyce, R. R., Allen, L. E. et al. 2012, \apj, 754, 19

\bibitem[Maund et al.(2013)]{mau13}
Maund, J. R., Spyromilio, J., H\"oflich, P. A., et al. 2013 \mnras, 433, L20

\bibitem[Nagata(1990)]{nag90}
Nagata, T. 1990, \apj, 348, L13

\bibitem[Ohyama et al.(2002)]{ohy02}
Ohyama, Y., Taniguchi, Y., Iye, M., et al. 2002, \pasj, 54, 891

\bibitem[Patat et al.(2009)]{pat09}
Patat, F., Baade, D., H\"oflich, P, et al. 2009, \aap, 508, 229

\bibitem[Patat et al.(2014)]{pat14}
Patat, F., Taubenberger, S., Cox, N. L. J., et al. 2014, \aap, submitted (arXiv: 1407.0136)

\bibitem[Pereira et al.(2013)]{per13} 
Pereira, R., Thomas, R. C., Aldering, G., et al. 2013, \aap, 554, A27

\bibitem[Phillips et al.(1999)]{phi99}
Phillips, M. M., Lira, P., Suntzeff, N. B., et al. 1999, \aj, 118, 1766

\bibitem[Phillips et al.(2013)]{phi13}
Phillips, M. M., Simon J. D., Morrell, N., et al. 2013, \apj, 779, 38

\bibitem[Richmond \& Smith(2012)]{ric12}
Richmond, M. W., \& Smith, H. A. 2012, JAVSO, 40, 872

\bibitem[Sakai \& Madore(1999)]{sak99}
Sakai, S., \& Madore, B. F. 1999, \apj, 526, 599

\bibitem[Schlafly \& Finkbeiner(2011)]{sch11}
Schlafly, E. F., \& Finkbeiner, D. P. 2011, \apj, 737, 103 

\bibitem[Serkowski, Mathewson, \& Ford(1975)]{ser75}
Serkowski, K., Mathewson, D. S., \& Ford, V. L. 1975, \apj, 196, 261

\bibitem[Shappee \& Stanek(2011)]{sha11}
Shappee, B. J. \& Stanek, K. Z. 2011, \apj, 733, 124

\bibitem[Smith et al.(2014)]{smi14}
Smith, P. S., Williams, G. G., Smith, N., Milne, P. A., Jannuzi, B. T.,
\& Green, E. M. 2014, \apjl, submitted (arXiv: 1111, 6626)

\bibitem[Turnshek et al.(1990)]{tur90}
Turnshek, D. A., Bohlin, R. C., Williamson, R. L., II, et al. 1990, \aj, 99, 1243

\bibitem[Vrba, Coyne, \& Tapira(1981)]{vrb81}
Vrba, F. J., Coyne, G. V., \& Tapia, S. 1981, \apj, 243, 489

\bibitem[Wang et~al.(1996)]{wan96}
Wang, L., Wheeler, J. C., Li, Z., \& Clocchiatti, A. 1996, 
\apj, 467, 435

\bibitem[Wang, Wheeler, \&H\"oflich(1997)]{wan97}
Wang, L., Wheeler, J. C., \& H\"oflich, P. 1997, \apj, 476, 27

\bibitem[Wang et~al.(2003)]{wan03}
Wang, L., Baade, D., H\"oflich, P., et al. 2003, \apj, 591, 1110

\bibitem[Wang et~al.(2006)]{wan06}
Wang, L., Baade, D., H\"oflich, P., et al. 2006, \apj, 653, 490

\bibitem[Wang and Wheeler(2008)]{wan08}
Wang, L., \& Wheeler, J. C. 2008, \araa, 46, 433

\bibitem[Wang et al.(2008)]{wanx08}
Wang, X., Li, W., Filippenko, A. V., et al. 2008, \apj, 675, 626

\bibitem[Wang et al.(2009)]{wan09}
Wang, X., Filippenko, A. V., Ganeshalingam, M., et al. 2009, \apjl, 699, L139

\bibitem[Welty et al.(2014)]{wel14}
Welty, D. E., Ritchey, A. M., Dahlstrom, J. A., \& York, D. G.
2014, arXiv: 1404.2639
 
\bibitem[Whittet et al.(1992)]{whi92} Whittet, D. C. B.,
Martin, P. G., Hough, J. H., et al. 1992, \apj, 386, 562
 
\bibitem[Whittet(2003)]{whi03} Whittet, D. C. B. 2003,
Dust in the Galactic Environment (2nd ed.; Bristol; IOP)

\bibitem[Wilking et al.(1980)]{wil80} Wilking, B. A., Lebofsky, M. J.,
Kemp, J. C., Martin, P. G., \& Rieke, G. H. 1980, \apj, 235, 905
 
\bibitem[Yanagisawa et~al.(2008)]{yan08}
Yanagisawa, K., Okita, K., Shimizu, Y., et al. 2008, Proc. SPIE, 7014, 701437

\bibitem[Zelaya et al.(2013)]{zel13}
Zelaya, P., Quinn, J. R., Baae, D., et al. 2013, \aj, 145, 27

\bibitem[Zheng et al.(2014)]{zhe14}
Zheng, W., Shivvers, I., Filippenko, A. V., et al. 2014, \apjl, 783, L24

\end{thebibliography}
\end{document}